# Influence of oxygen partial pressure on structural, transport and magnetic properties of Co doped TiO$_2$ films

[1]Bakhtyar Ali, [1]Abdul K Rumaiz, [1,2]S. Ismat Shah, [1]Arif Ozbay, [1]Edmund R Nowak

[1]Department of Physics and Astronomy, University of Delaware, Newark DE, 19716

[2]Material Science and Engineering, University of Delaware, Newark DE, 19716

## Abstract

Thin films of Co-TiO$_2$ are deposited on silicon and quartz substrates using Pulse Laser Deposition (PLD) process at various oxygen partial pressures ranging from 6.6 x 10$^{-3}$ Pascals (Pa) to 53 Pa. Crystal structure, transport and magnetic properties of reduced Co$_x$Ti$_{1-x}$O$_2$ (0 ≤ x ≤ 0.03) thin films are investigated and are found to have a strong dependence on the oxygen partial pressure. X-ray diffraction (XRD) data reveals the presence of mixed phase material containing both anatase and rutile. However, these phases intertransform with the change in the oxygen partial pressure in the chamber during the growth of the films. X-ray Photoelectron Spectroscopy (XPS) shows no Co or CoO related peaks for samples with Co concentration up to x=0.03. However, the oxygen 1s peaks are asymmetric suggesting the presence of oxygen vacancies. The transport and magnetic measurements show a clear dependence on the concentration of oxygen vacancies. There is an enhancement in the electrical conductivity and the magnetization as more vacancies are created in the material. The resistivity as a function of temperature ρ(T) follows the polaronic behavior and the activation energies obtained, ~100 to 150meV, are within the range that is typical for semiconducting materials.

# Introduction

In the rapidly evolving area of spintronics, the need for dilute magnetic semiconductors is increasing rapidly. Dilute magnetic semiconductors (DMS) is the class of materials where the magnetic impurities are lightly doped in the host semiconductor without forming a second phase. The most important condition for a DMS is that there must be spin polarization in the majority carrier band induced by exchange interaction between the magnetic dopant spins and the carriers [1]. Some examples of these types of materials are, Mn doped GaAs (III-V), Cr doped ZnTe, Mn doped Si, Co doped ZnO, Co/Cr doped $TiO_2$, etc. Among various DMS, Co-doped $TiO_2$ has attracted more attention because of the recent discovery of room temperature ferromagnetism in this system [2]. However, the origin of ferromagnetism is still controversial [3]. One of the key questions about the nature of the ferromagnetism manifested by these wide band gap DMS (e.g., doped $TiO_2$, ZnO, etc) is whether they are intrinsically ferromagnetic or if their magnetic behavior is the result of the precipitation of the magnetic impurities. Regarding the intrinsic nature of the magnetism, the work of Shinde et. al. [4] shows that the solubility limit of Co in $TiO_2$ is up to 0.02 beyond which precipitation occurs. However, upon annealing at elevated temperatures, uniform substitution of Co throughout the matrix of $TiO_2$ is observed. This proves that the phenomenon of magnetism is intrinsic and is not due to the clustering of Co metal. Similarly, the Co K-shell near-edge and extended X-ray absorption fine structure investigations of the charge state and local structure of Co in epitaxial Co doped $TiO_2$ films also ruled out the presence of metallic Co or its oxides [5]. Griffin et. al. also showed that there is no clustering of the magnetic impurity (Co). However, they explain the origin of magnetism in a slightly different way, by combining ferromagnetism with the dielectric properties of the material [6]. Although the work of Griffin et. al. excludes the carrier mediated ferromagnetism, there are some reports that show that the ferromagnetism in DMS is in fact carrier mediated [7, 8]. Various theories have been proposed for the interaction and magnetic coupling of the magnetic impurities in DMS. RKKY [7] interactions, Zener double exchange mechanism [8], polaron percolation model [9-13], etc., have been suggested to explain the observed magnetic and electrical transport features. Less attention paid, at least experimentally, has been directed at the role of oxygen vacancies which can affect the crystal structure of $TiO_2$ and, consequently, other physical properties, such as the electrical transport and magnetic properties in for example, Co doped and/or undoped $TiO_2$.

Oxygen vacancies have a significant effect on the physical and chemical properties of transition metal oxides [14, 15]. In case of many transition metal oxides, the oxygen vacancies play the role of electron traps and give rise to the donor levels in these materials [16]. $TiO_2$ is very sensitive to oxygen and can be reduced very easily when exposed to an oxygen deficient environment. Also, the requirement of the charge neutrality for the Co doped $TiO_2$ suggests that doping should lead the system to being reduced and that there should be some oxygen vacant sites [17-19]. Indeed some authors use this argument as the explanation for the magnetic and transport features observed in Co doped $TiO_2$ [20].

To elucidate the role of role of oxygen vacancies in affecting the structural, transport and magnetic properties of $TiO_2$, we studied undoped $TiO_2$ and Co-doped $TiO_2$ with various oxygen vacancies concentrations. Oxygen vacancies can be controlled in the samples by controlling the oxygen partial pressure in the PLD system during the film growth. In this paper we report on how the magnetic and electrical transport properties of the Co doped $TiO_2$ samples evolve with the concentration of oxygen vacancies. we discuss the progressive transition from relatively insulating to conducting behavior of the films as a function of the vacancy concentration in the films. A simple polaronic model is presented to describe the electrical transport of the films. Our results indicate that it is possible to control the concentration of vacancies in materials like $TiO_2$ and, thereby, tune the magnetic and transport properties.

## Experimental

The films studied in this work were prepared by a pulse laser deposition (PLD) technique. Targets with different concentrations of Co in $Ti_{1-x}Co_xO_2$, viz. x=0.00 (undoped $TiO_2$) 0.01 and 0.03 were prepared by solid state route. The undoped solid target was prepared by cold pressing high purity (99.999% Sigma-Aldrich) powder of titanium dioxide which was subsequently fired for several hours in air at 700ºC. The doped targets were prepared by mixing the appropriate amounts of Co (99.998% Alfa-Aesar) and TiO$_2$ powders. After the initial mixing, these powders were palletized and were fired at 700°C in an inert atmosphere of argon gas for 7 hours. These pallets were fired three more times with grindings between each firing so as to ensure the uniform mixing of Co and TiO$_2$.

The targets were laser ablated for 90 minutes by using an excimer laser (248 nm) at 400mJ energy with a repetition rate of 15 Hz. To avoid the exposure of one spot to the laser beam the target was constantly rotated by a stepper motor during the ablation. The base pressure in the chamber was ~1.3x 10$^{-5}$ Pascal. A mixture of oxygen and argon gas was used during the deposition. The oxygen partial pressures used were 6.6x 10$^{-5}$Pa, 0.013Pa, 0.066Pa, 0.133Pa, 0.66Pa, 1.33Pa, 13.33Pa.and 53Pa. Different oxygen partial pressures led to different oxygen contents in the samples. The growth rate of the films was 0.26 microns per minute. The films were initially characterized using X-ray Diffraction (XRD) and X-ray Photoelectron Spectroscopy (XPS). The thicknesses of the films were confirmed by both profilometer (DekTek) and the cross-sectional images were obtained from the Secondary Electron Microscopy (SEM). The transport and magnetic data were collected using Quantum Design's Physical Properties Measurement System (PPMS) and Quantum Design's Superconducting Quantum Interface Device (SQUID). These experiments were repeated several times to reproduce the results.

## Results and discussion

Initial structural characterizations were carried out using X-ray diffraction. The XRD patterns of the undoped TiO$_2$ and Co doped TiO$_2$ (x = 0.03) are shown in Fig. 1. All peaks are associated with the anatase or rutile phases, depending on oxygen partial pressure in the chamber during the synthesis. The only exception was the sample that was deposited under 1.33x10$^{-5}$Pa of oxygen partial pressure. In this case the material was reduced to such a high degree that it no longer remained TiO$_2$ but rather formed some other type of Ti-oxide. No peaks corresponding to Co or CoO phases were present. It is important to note that for a constant composition of Co, films in varying conditions of oxygen partial pressure were deposited. These changing environments of oxygen partial pressure resulted in the transformation of the crystal structure between anatase and rutile. Fig. 1a shows the XRD data for the pure $TiO_2$ samples that are deposited under various oxygen partial pressures. The XRD data shows that the samples that are grown under lower oxygen partial pressure have a dominant rutile (110) peak. This peak is suppressed for the samples that are grown under relatively high oxygen partial pressure where the anatase (101) peak grows gradually. This trend continues until the anatase phase becomes dominant. Thus, for pure $TiO_2$ samples the phase transformation occurs from rutile to anatase as the oxygen partial pressure is increased in the chamber during the deposition. A similar trend is observed in $Ti_{1-x}Co_xO_2$ samples for x = 0.01 and x = 0.02. However for x = 0.03 the transformation takes place in reverse order. That is, at lower oxygen partial pressure, films have predominantly anatase phase, in contrast to pure samples where rutile phase was observed for the films grown in lower oxygen partial pressure. This anatase phase transforms to rutile simply because of the increase in oxygen partial pressure at the time of deposition. Finally, at higher pressure (13.3 Pa), this trend stops and the rutile phase becomes the dominant phase. This behavior is shown in Fig. 1b. This transformation has a profound effect on the transport and magnetic properties of the material as will be discussed below.

X-ray photoelectron spectroscopy (XPS) was performed on all of the samples grown at various oxygen partial pressures ($6.6 \times 10^{-3}$ Pa to 53 Pa) using SSI-M-probe equipped with Al $K_\alpha$ monochromatic X-ray source and energy resolution of ~0.1eV. The base pressure in the XPS chamber was ~$1.3 \times 10^{-7}$ Pa prior to the start of data collection. For charge neutralization, a 1eV e-beam was applied to the sample. The charge correction was done using the carbon 1s peak as the reference for which the binding energy is ~284.5 eV. All the peaks were fitted employing the least square fit procedure. The full width at half maximum (FWHM) in all the fits was less than 2eV. In order to check for metallic cobalt in the sample, high resolution XPS scans were performed on the cobalt region. No cobalt related peak was detected. It should be noted that the Co concentration in the sample was only 0.03 which makes the observation of the peak difficult. However, a very interesting feature concerning the asymmetry of the oxygen 1s (O 1s) core spectra is observed which arises because of the non-stoichiometry of oxygen in $TiO_2$. Fig. 2 shows the normalized O 1s peaks recoded for x = 0.01 in $Ti_{1-x}Co_xO_2$ sample. The data in Fig 2a is obtained from the sample grown under $6.6\times10^{-3}$ Pa oxygen partial pressure, whereas Fig. 2b and Fig. 2c show O 1s regions of films grown under 0.13 Pa and 19.99 Pa oxygen partial pressures, respectively. In all these data, there is a clear asymmetry in the oxygen 1s peak. The profile clearly indicates the overlap of two symmetric peaks. This asymmetric peak can be deconvoluted into two symmetric peaks which are nominally assigned as the lower binding energy component (LBEC) and the higher binding energy component (HBEC). It has been shown in the literature that the HBEC arises because of the loss of oxygen (increase in the oxygen vacancy concentration) [21, 22]. The development of HBEC due to the increase in the oxygen vacancies concentration, obviously, leads to the asymmetry of the main peak. Oxygen 1s X-ray photoemission spectrum shows that the asymmetry of O 1s peak is more pronounced in the samples grown in lower oxygen partial pressure as compare to those grown in higher oxygen partial pressure. Furthermore, the relative area under the curve (area of HBEC peak/area of LBEC) is determined to be ~0.47 for lower oxygen partial pressure (Fig. 2a) which is comparable to the area under the curve for LBEC. This large amount of oxygen deficiency could lead to the formation of some other oxide phase of titanium, for example, $Ti_2O_3$. This is in agreement with the XRD data where we do not observe the diffraction pattern for anatase or rutile $TiO_2$. On the other hand, for samples grown under higher oxygen partial pressure, shown in Fig. 2b and Fig. 2c, areas under the curve are comparatively smaller than that in Fig. 2a and are ~0.29 and 0.04, respectively. The relatively large contribution of the HBEC peak in the case of sample grown in lower oxygen partial pressure strongly suggests the presence of more oxygen vacancies. This is most likely because of the loss of oxygen atoms from the host material during the ablation of the target with high energy laser beam and the reduced oxygen partial pressure at elevated temperatures (~700°C) that results in the reduction of the $TiO_2$ and the formation of oxygen vacancies in the lattice. Moreover, the OH ions absorbed from the air during the transportation of the sample from the PLD chamber to XPS can also lead to the formation of the HBEC. However, in our case, if one looks at the data carefully one could see that there is a systematic increase in the HBEC peak intensity as we decrease the oxygen partial pressures. If it were due to the OH related peak, all samples would have had the same intensity peak. This suggests that the trend is indeed related to the O vacancies.

These oxygen vacancies are expected to generate free carriers (electrons) that can help in mediate exchange interaction effects between the magnetic impurities and also affect the electrical conduction in the material. Oxygen vacancies are indeed responsible for the enhancement in the electrical conductivity and magnetization of the material, as will be discussed in the following sections. The enhancement in both the electrical conduction and magnetization signifies the role of defects/vacancies in laser ablated $TiO_2$ and suggests that the magnetic properties in such a material are not due solely to magnetic impurities. These findings are consistent with the results obtained by Kaspars et. al. where they correlated the onset of ferromagnetism with the presence of extended structural defects [1, 23].

Electrical resistivity as a function of temperature was measured from room temperature to 5K. The measurements were carried out on films grown on quartz substrates under various oxygen partial pressures, ranging from $6.6 \times 10^{-3}$ Pa to 1.06 Pa. Sample resistivity was determined from either a conventional four-probe resistance measurement or using an electrometer, depending on the samples resistance. It is evident in Fig. 3 that films grown under lower oxygen partial pressure show more conductive behavior as compared to those which are grown under high oxygen partial pressure. The inset in Fig. 3 shows data for samples that were deposited under very low oxygen partial pressure. The resistivity vs temperature plot shows semiconducting behavior. Samples that were deposited at higher oxygen partial pressure, above ~0.13Pa, exhibit a sharp increase in resistivity with the decrease in temperature. As the temperature is decreased further, the resistivity becomes constant and remains so down to the lowest measurement temperature of ~2K. This upturn in the resistivity profile shifts towards higher temperatures for samples that are grown at higher oxygen partial pressures which presumably, have less concentration of oxygen vacancies. We define $T_p$ as the temperature that corresponds to the maximum resistivity. $T_p$ shifts towards higher temperatures as the concentration of oxygen vacancies is decreased. This trend is suggestive of a polaronic behavior in this material because the material has enough oxygen defects (vacancies) which possibly can trap electrons and produce polarons [24]. These are usually small or large polarons, depending on the effective mass of the trapped electron. Small polaron is hard to move due to its larger effective mass as compare to large polarons. This lowers the conductivity. However, small polaron can also lead to higher mobility through percolation mechanism where the conduction occurs in the percolation channels. These channels come into existence when the concentration of the oxygen vacancies reaches the percolation threshold [24-26]. This mechanism seems very likely in our system because when the sample is deposited under relatively rich oxygen environments which eliminate the oxygen vacancies, lower conductivity is observed. This indicates that polaron formation is suppressed by the oxygen rich environments during the deposition, i.e., the number of small polarons is below the percolation threshold.

We fit our transport data to the polaronic model, where the nearest neighbor of small polaron leads to mobility with a thermally activated form. The resistivity is given by

$$\ln \rho = \ln A + \frac{W}{k_B T} \qquad (1)$$

$$\text{Where } W = \frac{E_p}{2} - t \qquad (2)$$

A is a temperature independent prefactor, W is the effective activation energy, $E_p$ is the polaron formation energy and t is the transfer integral [27]. We note that only the data collected from the samples that are prepared in the oxygen partial pressure ~ 0.26Pa or below, are well described by the polaronic model. The apparent reason seems to be that the start of the upturn in the resistivity (as discussed above) for higher $O_2$ partial pressure deposited samples occurs at relatively higher temperature. At higher temperature the resistivity values are comparatively smaller. We infer from this that with less number of vacancies in the material, the polarons find themselves more confined and need more energy for their conduction.

Linear fits of the data (0.26 Pa samples and below), yielded activation energies of ~147meV for the polaron hopping mechanism. These values are in a good agreement with those reported for other semiconducting oxides [28]. Fig. 4 shows one of the linear fits to the data for x=0.03 Co doped sample which is deposited under 0.06Pa oxygen partial pressure. It can also be seen that the fit of the data to this model extends only to the region before the large upturn (i.e. for $T>T_p$) in resistivity. Below $T_p$ $\rho$ is nearly temperature independent. A plausible explanation is that polarons freeze out near $T=T_p$ and transport is dominated by a parallel conductive path which is metallic like.

Room temperature magnetic hysteresis loops for x=0.01 and 0.03 are shown in Fig. 5 and Fig. 6, respectively. In all these magnetic data, the loops (a) are taken for a sample that was grown in reduced oxygen atmosphere (~6.6x10$^{-3}$ Pa) whereas data of the loops (b) are collected on a sample which was grown under a relatively high oxygen partial pressure (0.66 Pa). The data for x=0.01 sample show a slightly decreasing trend of magnetization as the field is swept to higher values. This is because the sample's ferromagnetic signal is weaker as compare to the diamagnetic signal from the substrate. In this case the diamagnetic contribution from the substrate dominates the magnetization of the material. This kind of behavior is not seen in the x=0.03 sample because the magnetization value of the sample is large enough that the substrate diamagnetic response to the higher applied field (~1T) is suppressed. The same diamagnetic signal is also observed in the case of pure $TiO_2$ (undoped) but in this case there is no ferromagnetic signal detected. On the other hand, saturation magnetization ($M_s$) increases for the samples which are grown in lower oxygen partial pressure, almost by three fold, as compare to those sample of the same Co concentration which are grown in relatively high oxygen partial pressure (0.66Pa). This probably is due to the increase in the number of the oxygen vacancies which leads to an obvious increase in the magnetization, as is seen from the magnetization data in Fig. 5 and Fig. 6.

All the above findings were confirmed by reproducing them several times with samples of various cobalt concentration viz. x = 0.01, 0.02, 0.03, 0.06, etc. From these observations we draw one important conclusion that the observed behavior of magnetization enhancement has not been observed in the case of Co metal. This means that certainly some amount of the Co, if not all, has been distributed in $TiO_2$ lattice as a substituted magnetic impurity. Thus this eliminates the possibility that the observed ferromagnetism is due to the Co precipitation, rather the dominant role in operating the ferromagnetism is that of the combination of defects induced by the oxygen vacancies and magnetic impurity in the form of cobalt.

The underlying mechanism for this enhancement of magnetization and conductivity lies in the spin-split donor impurity-band model, proposed by Coey etal [29]. According to this model when there is an overlap of the donor-impurity band, which is usually spin spilt, with the 3d band of the transition metal, high Curie temperatures ($T_c$) are observed. Thus, we believe that there is some overlap between the magnetic impurities (Co, in this case) doped in $TiO_2$ and the states arisen from the defects. The origin of the donor impurity band most likely comes from the oxygen defects which, in turn, are the outcome of the lower oxygen partial pressure and the high energy laser ablation of the target. These defects, such as oxygen vacancies, are capable of trapping between one or two electrons, and are known as F-centers [30, 31]. In fact, we have observed these donor states in our high resolution valence band XPS spectra [32]. The value of $M_s$ is 0.283emu/cc for the x=0.03 sample grown in $6.6 \times 10^{-3}$ Pa oxygen partial pressure whereas $M_s$ is 0.111emu/cc for the sample with the same Co concentration in $TiO_2$ which is grown at a higher partial pressure of oxygen of 0.66Pa. This trend is consistent with the resistivity data which show less resistivity when a lower oxygen partial pressure is maintained during the deposition of the material and vice versa. It seems plausible that due to the magnetic ordering, the chances of scattering are reduced, which result in lower values of resistivity. The magnetic order is related to the vacancies content in the sample. Once magnetic order is reached, the scattering centers are decreased, and more conduction occurs.

## Conclusion

Cobalt doped $TiO_2$ samples deposited on Si and quartz substrates were studied in the context of oxygen vacancy concentrations and the cobalt doping levels. The O-1s peaks revealed qualitatively that the samples were oxygen deficient. The number of the oxygen vacancies is higher in the samples deposited under a reduced oxygen atmosphere. Crystalline structural transformations take place as a function of the oxygen partial pressure during the films deposition. These transformations occur from rutile to anatase for the nominal cobalt concentration, less than or equal to 0.02, when the oxygen partial pressure is increased in PLD chamber. The phase transformation depended upon the Co concentration. For nominal Co concentration above 0.02, the transformation took place in a reverse order. That is, at lower oxygen partial pressure the dominant phase was now anatase which transformed to rutile as the pressure is increased during deposition. The concentration of oxygen vacancies also has a profound effect on the magnetic and transport behavior of the material. The resistivity become smaller as the concentration of vacancies is increased in the material by providing the chamber with a lower oxygen partial pressure as the material is deposited. Concomitantly, the magnetization increases with increased vacancy concentration. This increase in the magnetization was consistent with the resistivity data. The resistivity data suggests that the conduction mechanism is predominantly polaronic when the films were grown in oxygen partial pressure less than or equal to 0.26 Pa. The enhancement in both the conduction and magnetization highlights the role of defects/predominantly oxygen vacancies in this material.

**Figures captions:**

**Figure-1** (a) XRD patterns for pure $TiO_2$ (undoped) obtained for the samples grown under different oxygen partial pressures. (b) XRD patterns for 3% Co doped samples grown under different oxygen partial pressures.

**Figure 2:** XPS spectra of O 1s core level obtained from the 1% Co doped $TiO_2$ samples that are grown (a) under oxygen partial pressure of $6.6 \times 10^{-3}$ Pa, (b) Oxygen partial pressure of 0.13 Pa and (c) 19.99 Pa of oxygen partial pressure.

**Figure3:** Resistivity as a function of temperature for the Co doped $TiO_2$ films (Co=3%) grown at different Oxygen partial pressures. Curves (a), (b), (c), (d) and (e) are ($\rho = \rho_0 T \exp(E_a/k_B T)$) respectively deposited under the oxygen partial pressure of $6.6 \times 10^{-3}$ Pa, 0.06 Pa, 0.13 Pa, 0.26 Pa and 1.06 Pa. Inset shows the plot of the two flat curves (a) and (b) in the main figure.

**Figure4:** Fitting of the resistivity data to the Polaronic Model for the 3% Co doped sample grown under oxygen partial pressure 0.06 Pa. The fit is for the temperature range 300>T>142 (Kelvin).

**Figure5:** Magnetic hysteresis loops for 1% Co doped $TiO_2$ samples grown in **(a)** $6.6 \times 10^{-3}$ Pa oxygen partial pressure and **(b)** in 0.66 Pa oxygen partial pressure.

**Figure6:** Magnetic hysteresis loop for the 3% Co doped $TiO_2$ for the samples grown (a) in $6.6 \times 10^{-3}$ Pa of oxygen partial pressure and (b) 0.66 Pa of oxygen partial pressure.

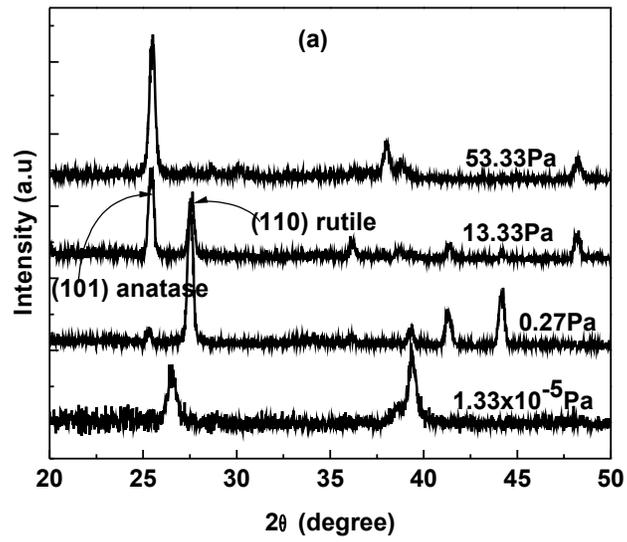

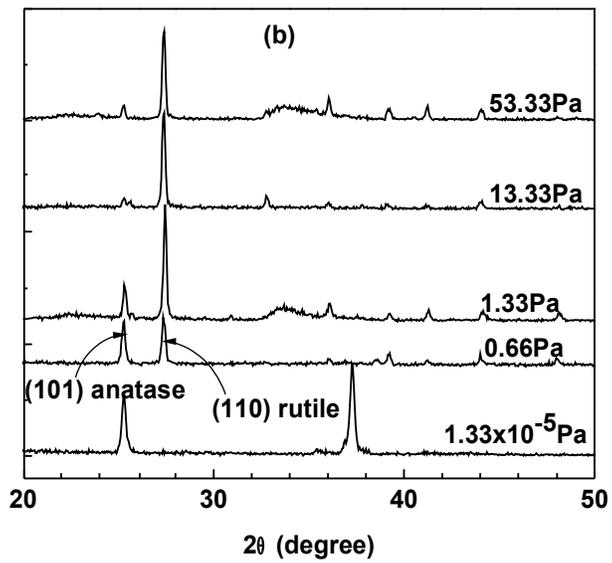

Figure-1 Bakhtyar Ali, Abdul K Rumaiz, S. Ismat Shah, Arif Ozbay, Edmund R Nowak

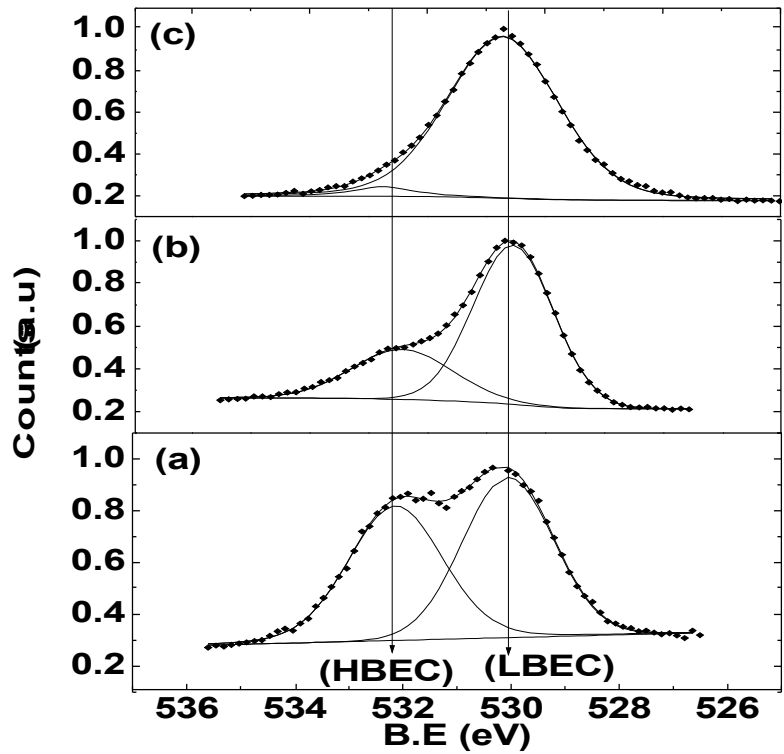

Figure-2 Bakhtyar Ali, Abdul K Rumaiz, S. Ismat Shah, Arif Ozbay, Edmund R Nowak

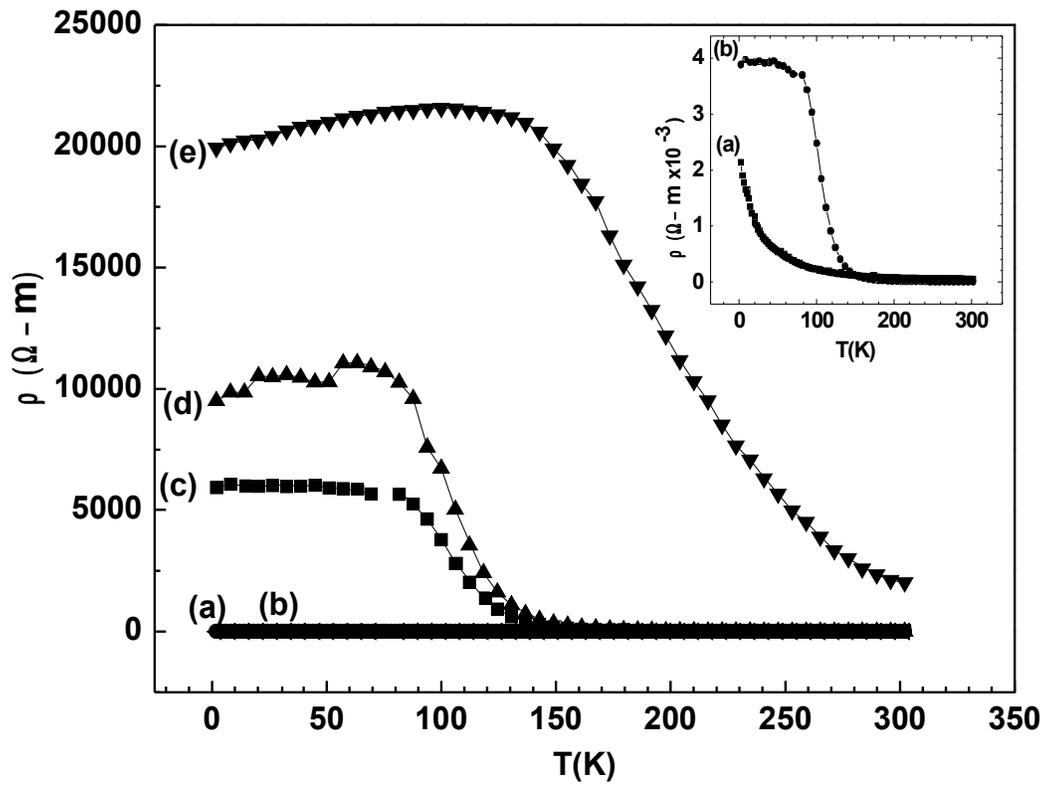

Figure-3 Bakhtyar Ali, Abdul K Rumaiz, S. Ismat Shah, Arif Ozbay, Edmund R Nowak

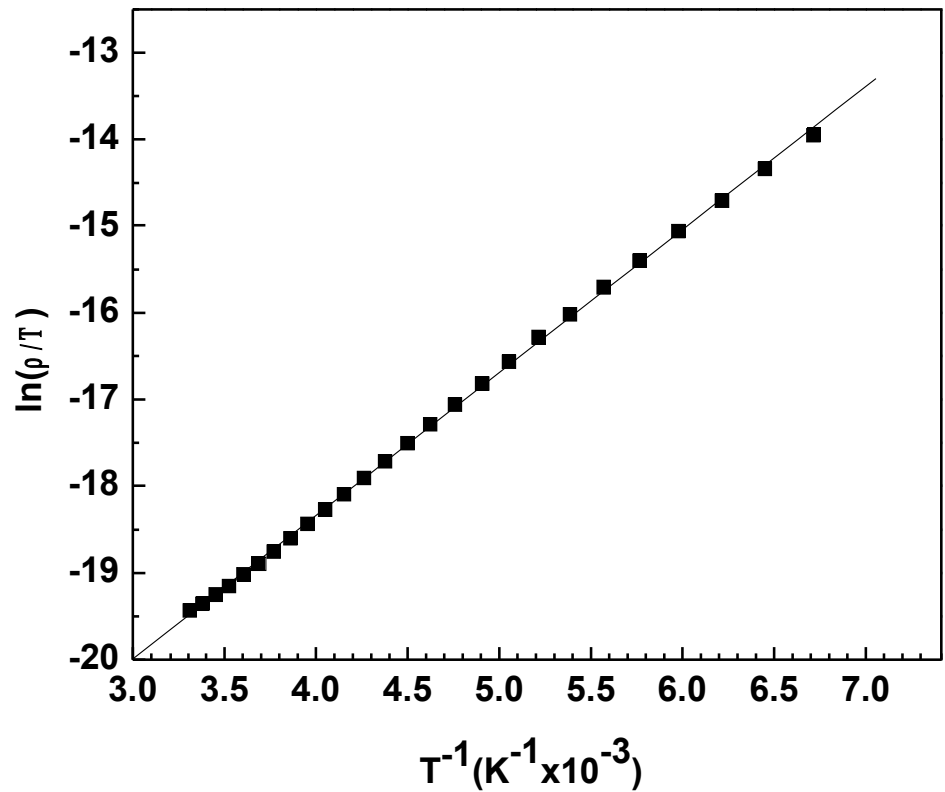

Figure-4 Bakhtyar Ali, Abdul K Rumaiz, S. Ismat Shah, Arif Ozbay, Edmund R Nowak

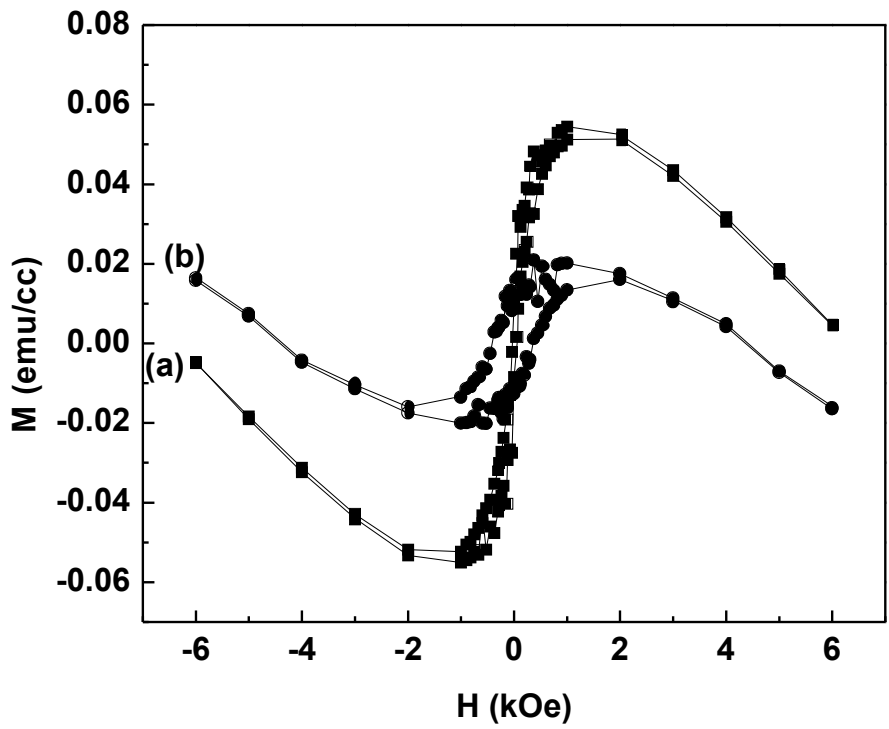

Figure-5 Bakhtyar Ali, Abdul K Rumaiz, S. Ismat Shah, Arif Ozbay, Edmund R Nowak

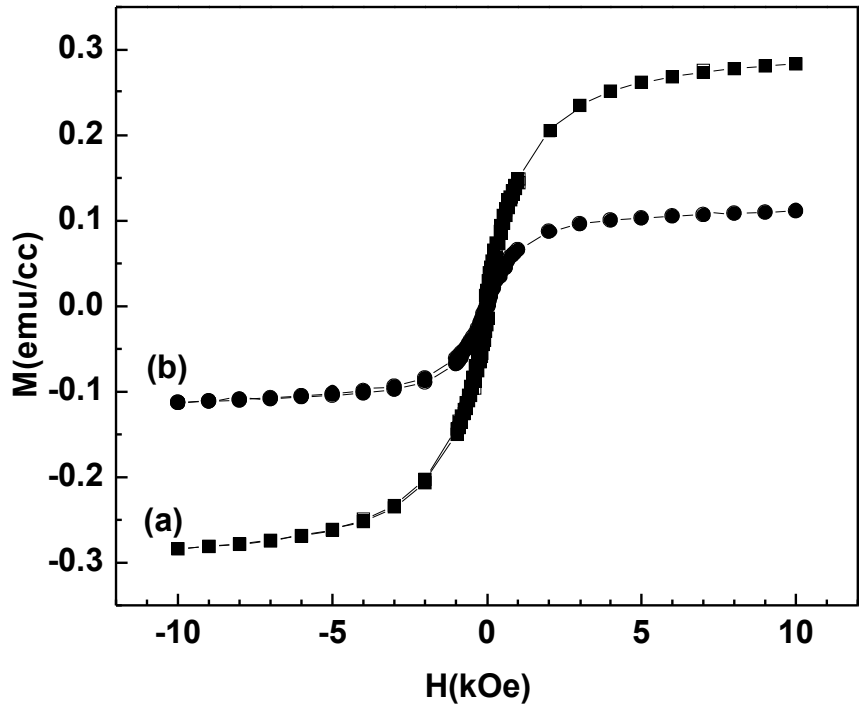

Figure-6 Bakhtyar Ali, Abdul K Rumaiz, S. Ismat Shah, Arif Ozbay, Edmund R Nowak